# Probing the Validity of EFT Approaches in Higgs + Jet Production at $\sqrt{s}$ = 13 TeV

Shubham Yadav[1][0009-0001-3560-6665] and Prashant Shukla[1][0000-0003-1456-6593]

[1]Dr. Hari Singh Gour Vishwavidyalaya, Sagar, India - 470003.
shubham.hep@gmail.com

**Abstract.** The production of a Higgs boson with a jet is one of the most important processes studied at the LHC. It serves as a sensitive probe of high-energy dynamics and provides a powerful testing ground for the Standard Model, while also offering sensitivity to possible new physics effects.
In this work, we perform a phenomenological study of Higgs + jet production in proton-proton collisions at 13 TeV center of mass energy, using the MadGraph5_aMC@NLO framework. We compare predictions obtained from the full loop induced Standard Model with those derived within the HEFT and SMEFT frameworks. Taking the Higgs transverse momentum as a probe, we study normalized distributions and ratio spectra to test the effective field theories. Our results show that effective approaches successfully reproduce the Standard Model behavior in the low transverse-momentum region. However, sizable deviations emerge at high $p_T$, signaling the breakdown of the effective approximation. These findings highlight the importance of carefully accounting for kinematic effects when applying EFT-based methods to precision Higgs studies and provide a quantitative estimate of their range of validity.

## 1 Introduction

The discovery of the Higgs boson at the Large Hadron Collider (LHC) has enabled precision studies of its production mechanisms and interactions, providing a powerful framework to test the Standard Model (SM) and probe possible effects of new physics [1,2]. Current measurements of Higgs production and decay rates are in good agreement with SM predictions but still allows room for potential deviations arising from physics beyond the Standard Model (BSM) [3].
The dominant production mechanism of the Higgs boson at the LHC is gluon fusion, which proceeds through a heavy-quark loop, primarily mediated by the top quark [4]. As a result, the Higgs-gluon interaction is loop-induced and sensitive to contributions from heavy states. Precise characterization of this coupling is therefore essential for understanding possible modifications due to new physics.
In practical calculations, the loop-induced Higgs-gluon coupling is often approximated using an effective field theory approach. In the limit of an infinitely heavy top quark, the interaction reduces to a local dimension-5 operator, forming the basis of the HEFT.

This approximation has proven to be highly successful in describing inclusive observables and total cross sections. However, it is expected to lose validity when the energy scale of the process becomes comparable to the top-quark mass [4,5].
At large transverse momentum of the Higgs boson or the associated jet, the internal structure of the loop becomes resolvable, leading to deviations from the point-like approximation inherent in heft. Consequently, differential distributions in the high-$p_T$ regime offer a powerful way to test the validity of effective descriptions.
A systematic framework to capture such deviations is provided by effective field theory extensions that include higher-dimensional operators. While the leading dimension-5 operator describes the dominant interaction in the heavy-top limit. These operators can modify both the strength and kinematic behavior of the Higgs-gluon coupling, thereby affecting observables such as $p_T$ distributions [6,7].
Standard Model Effective Field Theory (SMEFT), have been developed to study deviations in Higgs interactions. However, many studies focus primarily on inclusive observables or modifications of existing operators, while the role of higher-dimensional gluonic operators becomes increasingly important in boosted regimes.
In this work, we investigate the validity of effective field theory descriptions of the Higgs-gluon interaction by studying Higgs boson production in association with a jet at the LHC with $\sqrt{s}$ = 13 TeV. We compare predictions from the HEFT approximation with full loop-induced calculations and examine the impact of higher-dimensional operators within the SMEFT framework. Particular emphasis is placed on the transverse momentum distributions of the Higgs boson, which provide a sensitive probe of finite top-mass effects and potential new physics contributions.

## 2 Simulation and Analysis Framework

In this study we have used MadGraph5_aMC@NLO event generator which provides computations of cross sections, the generation of hard events and their matching with event generators, and the use of a variety of tools relevant to event manipulation and analysis for SM and BSM Phenomenology [8]. We simulate the process p p → h j with di-photon decay channel, for 400k events. In table1, we have shown the detailed info about parton level cuts.

**Table1.** Parton level cuts definition

| Parameter | Value | Explanation |
|---|---|---|
| ptj | 20.0 | Minimum $p_T$ for jets |
| pta | 10.0 | Minimum $p_T$ for photons |
| etaj | 5.0 | Maximum jet rapidity |
| etaa | 2.5 | Maximum photon rapidity |
| drjj | 0.4 | Minimum distance between jets |
| draa | 0.4 | Minimum distance between photons |

## 3 Results and Discussion

We have shown the Higgs $p_T$ obtained from HEFT and loop_sm predictions and the ratio panel indicates that HEFT agrees with the full loop calculation within approximately 5-10% for $p_T \lesssim 300$ GeV. However, as the transverse momentum increases, deviations become more pronounced. (See Fig. 1)

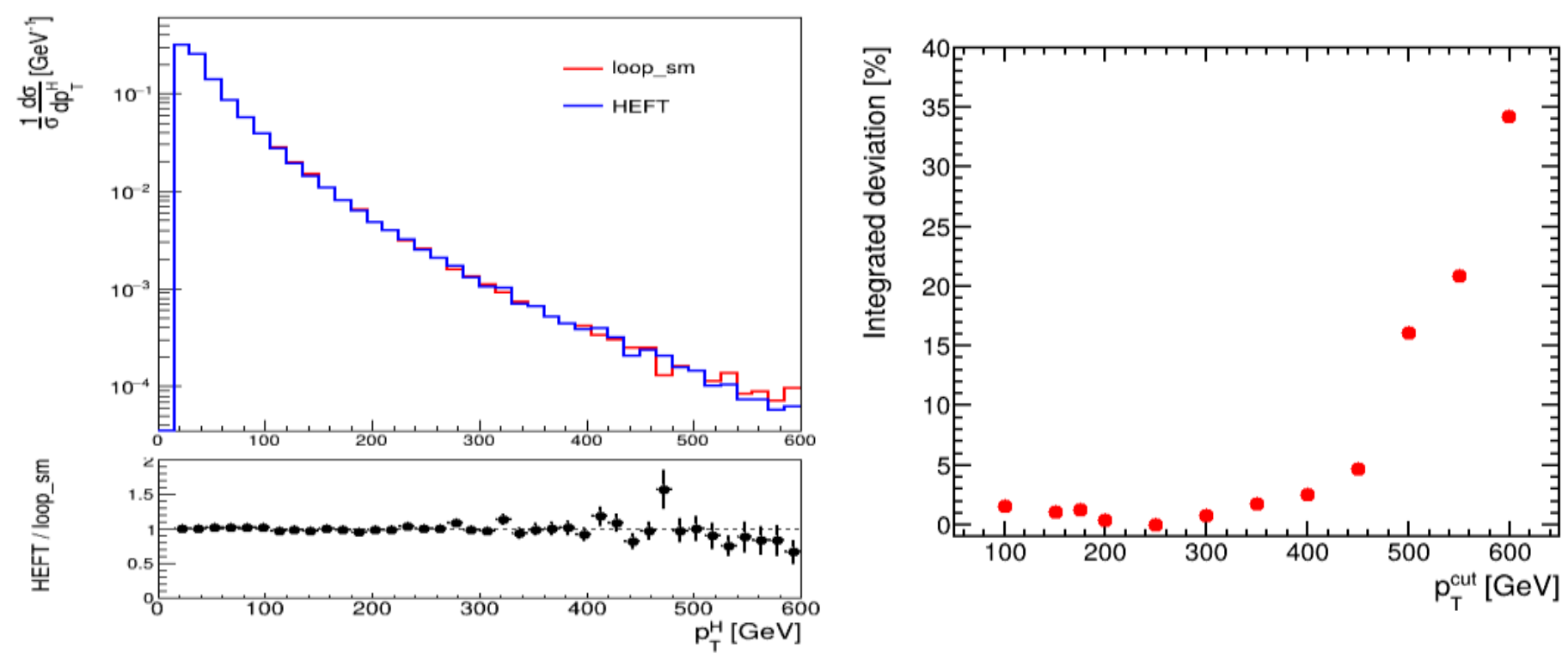

**Fig. 1.** Higgs transverse momentum spectra (Left) and Integrated deviation of Higgs transverse momentum (Right)

In the boosted regime ($p_T \gtrsim 400$ GeV), the discrepancy grows significantly, reaching 20-30% at high $p_T$. This behavior reflects the breakdown of the heavy top quark approximation, as the internal structure of the loop becomes resolved at high energy scales.

To further quantify the difference between HEFT and the full SM prediction, the integrated deviation as a function of the transverse momentum cut is shown in Fig. 1 (Right). The deviation remains small (< 5%) up to $p_T \sim 350$ GeV, indicating that HEFT provides a reliable approximation in this region. Beyond this scale, the deviation increases rapidly. This clearly demonstrates that high-$p_T$ observables are highly sensitive to finite Top-mass effects and provide a stringent test of EFT validity.

The impact of higher-dimensional operators is studied by comparing SMEFT predictions with different values of the Wilson coefficient cHG to the full SM result, as shown in Fig. 2.

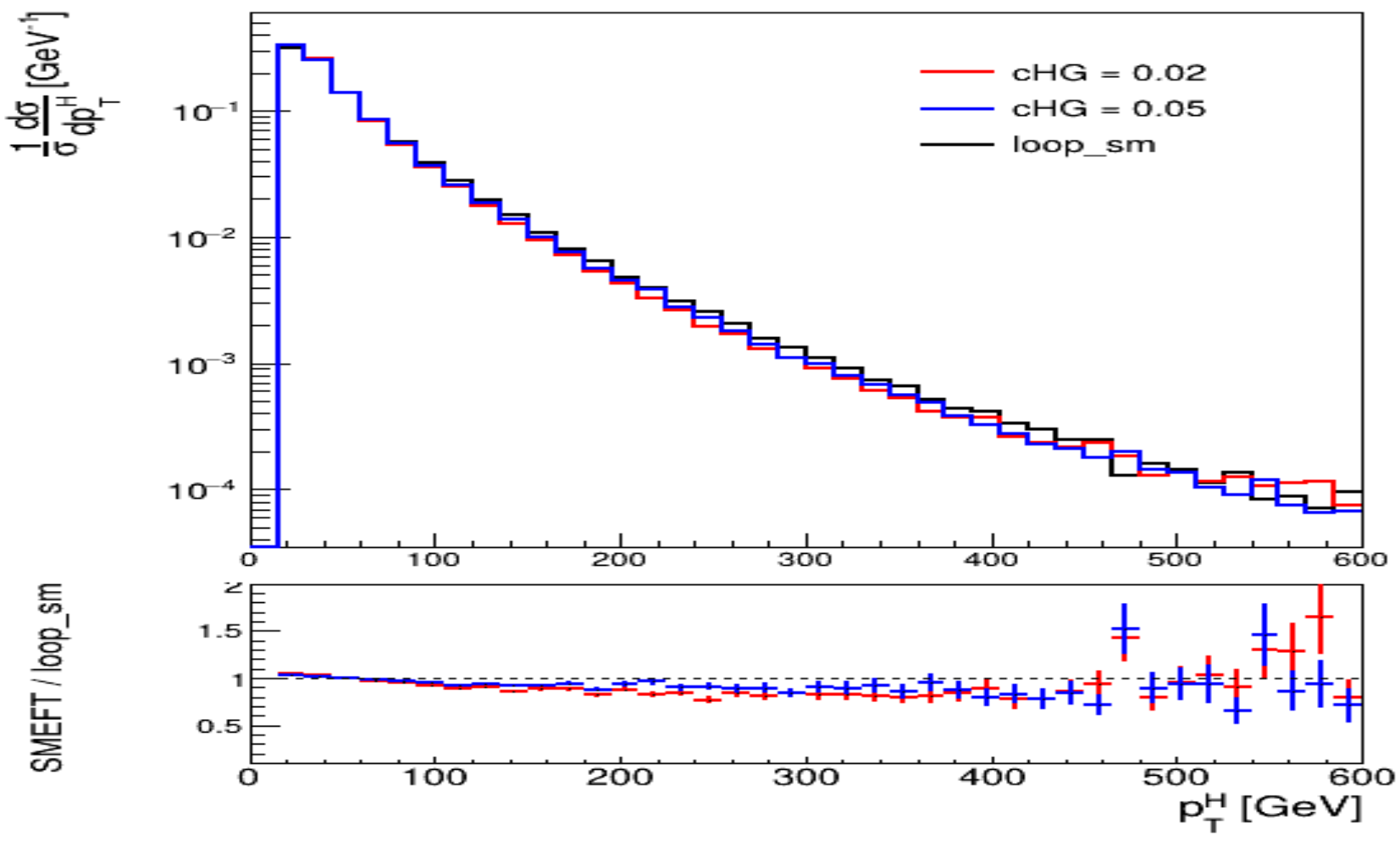

**Fig. 2.** Higgs transverse momentum spectra with SMEFT predictions.

At low transverse momentum, all distributions closely follow the SM prediction, indicating limited sensitivity to new physics in this region. However, in the high-$p_T$ regime, clear deviations emerge. The ratio plots show that the deviations increase with $p_T$. and for larger values of cHG lead to stronger distortions, effects become significant for $p_T \gtrsim 350$-400 GeV. This demonstrates that boosted Higgs production provides a powerful probe of higher-dimensional operators and potential new physics effects.

## 4 Summary and Outlook

In this work, we compared Higgs $p_T$ predictions from HEFT, the full loop calculation, and SMEFT. HEFT agrees well with the full Standard Model at low $p_T$, but starts to break down beyond $\sim$ 300-400 GeV, where deviations grow significantly. This shows that finite top-mass effects become important in the boosted region. The integrated deviation confirms that HEFT is reliable only up to about 350 GeV.
From the SMEFT study, we see that new physics effects are small at low $p_T$ but become clearly visible at high $p_T$, especially for larger values of cHG. This makes boosted Higgs production a sensitive probe of higher-dimensional operators.
Future improvements could include detector-level studies, higher-order corrections, and a more complete SMEFT analysis. Comparing with data from the LHC will be crucial to test these results.